# ANALYSIS OF COMPARATIVE DATA WITH HIERARCHICAL AUTOCORRELATION

By Cécile Ané

*University of Wisconsin—Madison*

The asymptotic behavior of estimates and information criteria in linear models are studied in the context of hierarchically correlated sampling units. The work is motivated by biological data collected on species where autocorrelation is based on the species' genealogical tree. Hierarchical autocorrelation is also found in many other kinds of data, such as from microarray experiments or human languages. Similar correlation also arises in ANOVA models with nested effects. I show that the best linear unbiased estimators are almost surely convergent but may not be consistent for some parameters such as the intercept and lineage effects, in the context of Brownian motion evolution on the genealogical tree. For the purpose of model selection I show that the usual BIC does not provide an appropriate approximation to the posterior probability of a model. To correct for this, an effective sample size is introduced for parameters that are inconsistently estimated. For biological studies, this work implies that tree-aware sampling design is desirable; adding more sampling units may not help ancestral reconstruction and only strong lineage effects may be detected with high power.

**1. Introduction.** In many ecological or evolutionary studies, scientists collect "comparative" data across biological species. It has long been recognized [Felsenstein (1985)] that sampling units cannot be considered independent in this setting. The reason is that closely related species are expected to have similar characteristics, while a greater variability is expected among distantly related species. "Comparative methods" accounting for ancestry relationships were first developed and published in evolutionary biology journals [Harvey and Pagel (1991)], and are now being used in various other fields. Indeed, hierarchical dependence structures of inherited traits arise in many areas, such as when sampling units are genes in a gene family [Gu (2004)], HIV virus samples [Bhattacharya et al. (2007)], human cultures [Mace and Holden (2005)] or languages [Pagel, Atkinson and Meade (2007)]. Such tree-structured units show strong correlation, in some way similar to the correlation encountered in spatial statistics. Under the spatial "infill" asymptotic where a region of space is filled in with densely sampled points, it is known that some parameters are not consistently estimated [Zhang and Zimmerman (2005)]. It is shown here that inconsistency is also the fate of some parameters under hierarchical dependency. While spatial statistics is now a well recognized field, the statistical analysis of tree-structured data has been mostly developed by biologists so far. This paper deals with a classical regression framework used to analyze data from hierarchically related sampling units [Martins and Hansen (1997), Housworth, Martins and Lynch (2004), Garland, Bennett and Rezende (2005), Rohlf (2006)].

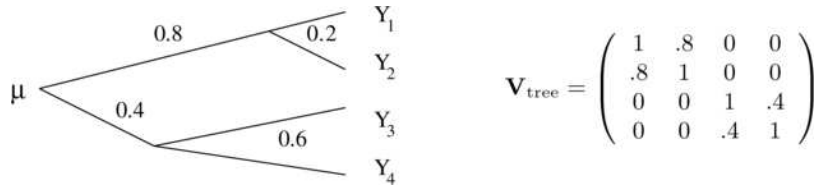

Fig. 1.  *Example of a genealogical tree from 4 units (left) and covariance matrix of vector* **Y** *under the Brownian motion model (right).*

*Hierarchical autocorrelation.*  Although species or genes in a gene family do not form an independent sample, their dependence structure derives from their shared ancestry. The genealogical relationships among the units of interest are given by a tree (e.g., Figure 1) whose branch lengths represent some measure of evolutionary time, most often chronological time. The root of the tree represents a common ancestor to all units considered in the sample. Methods for inferring this tree typically use abundant molecular data and are now extensively developed [Felsenstein (2004), Semple and Steel (2003)]. In this paper the genealogical tree relating the sampled units is assumed to be known without error.

The Brownian model (BM) of evolution states that characters evolve on the tree with a Brownian motion (Figure 2). After time $t$ of evolution, the character is normally distributed, centered at the ancestral value at time 0 and with variance proportional to $t$. Each internal node in the tree depicts a speciation event: an ancestral lineage splitting into two new lineages. The descendant lineages inherit the ancestral state just prior to speciation. Each lineage then evolves with an independent Brownian motion. The covariance matrix of the data at the $n$ tips $\mathbf{Y} = (Y_1, \ldots, Y_n)$ is then determined by the tree and its branch lengths:

$$\mathbf{Y} \sim \mathcal{N}(\mu, \sigma^2 \mathbf{V}_{\text{tree}}),$$

where $\mu$ is the character value at the root of the tree. Components of $\mathbf{V}_{\text{tree}}$ are the times of shared ancestry between tips, that is, $V_{ij}$ is the length shared by the paths from the root to the tips $i$ and $j$ (Figure 1). The same structural covariance matrix could actually be obtained under other models of evolution, such as Brownian motion with drift, evolution by Gaussian jumps at random times or stabilizing selection in a random environment [Hansen and Martins (1996)]. The i.i.d. model is obtained with a "star" tree, where all tips are directly connected to the root by edges of identical lengths. Another model of evolution assumes an Ornstein–Uhlenbeck (OU) process and accounts for stabilizing selection [Hansen (1997)]. The present paper covers the assumption of a BM structure of dependence, although several results also apply to OU and other models. As the Brownian motion is reversible, the tree can be re-rooted. When the root is moved to a new node in the tree, the ancestral state $\mu$ represents the state of the character at that new node, so re-rooting the tree corresponds to a re-parametrization.

*The linear model.* A frequent goal is to detect relationships between two or more characters or to estimate ancestral traits [Schluter et al. (1997), Pagel (1999), Garland and Ives (2000), Huelsenbeck and Bollback (2001), Blomberg, Garland and Ives (2003), Pagel, Meade and Barker (2004)]. In this paper I consider the linear model $\mathbf{Y} = \mathbf{X}\beta + \varepsilon$ with $\varepsilon \sim \mathcal{N}(0, \sigma^2 \mathbf{V}_{\text{tree}})$ as derived from a BM evolution on the tree. When the matrix of predictors $\mathbf{X}$ is of full rank $k$, it is well known that the best linear unbiased estimator for $\beta$ is

$$\hat{\beta} = (\mathbf{X}^t \mathbf{V}_{\text{tree}}^{-1} \mathbf{X})^{-1} \mathbf{X}^t \mathbf{V}_{\text{tree}}^{-1} \mathbf{Y}.$$

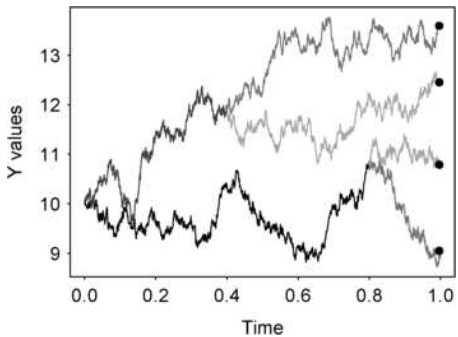

FIG. 2. *Simulation of BM evolution along the tree in Figure 1. Ancestral state was $\mu = 10$. Observed values of $Y$ are marked by points.*

Random covariates are typically assumed to evolve with a BM on the same tree as $\mathbf{Y}$. Fixed covariates are also frequently considered, such as determined by a subgroup of tips.

Although this model has already been used extensively, the present paper is the first one to address its asymptotic properties. For a meaningful asymptotic framework, it is assumed that the root of the tree is fixed while units are added to the sample. The reason is that the intercept relates to the ancestral state at the root of the tree. If the root is pushed back in time as tips are added to the tree, then the meaning of the intercept changes and there is no hope of consistency for the intercept. The assumption of a fixed root is just a rooting requirement. It does not prevent any unit to be sampled.

Asymptotic results assume the sample size goes to infinity. I argue here that this is relevant in real biological studies. For instance, studies on phylogenetically related viral samples have included hundreds of samples [Bhattacharya et al. (2007)]. Pagel, Atkinson and Meade (2007) have built and used a tree relating as many as 87 Indo-European languages. Many groups count an incredibly large number of species. For instance, there are about 20,000 orchid species to choose from [Dressler (1993)], over 10,000 species of birds [Jønsson and Fjeldså (2006)], or about 200 wild potato species [Spooner and Hijmans (2001)]. In addition, studies can consider subpopulations and even individuals within species, so long as they are related by a divergent tree.

*Organization.* The main results are illustrated on real examples in Section 2. It is shown that $\hat{\beta}$ is convergent almost surely and in $L^2$ norm in Section 3. In Section 4 then, I show that some components of $\hat{\beta}$ are *not* consistent, converging to some random value. This is typically the case of the intercept and of lineage effect estimators, while estimates of random covariate effects are consistent. I investigate a sampling strategy—unrealistic for most biological settings—where consistency can be achieved for the intercept in Section 4. With this sampling strategy, I show a phase transition for the rate of convergence: if branches are not sampled close to the root of the tree fast enough, the rate of convergence is slower than the usual $\sqrt{n}$ rate. In Section 5 I derive an appropriate formula for the Bayesian Information Criterion and introduce the concept of effective sample size. Applications to biological problems are discussed in Section 6, as well as applications to a broader context of hierarchical models such as ANOVA.

**2. Illustration of the main results.** Davis et al. (2007) analyzed flower size diameter from $n = 25$ species. Based on the plants' tree (Figure 3 left) assuming a simple BM motion with no shift, calculations yield an effective sample size $n_e = 5.54$ for the purpose of estimating flower diameter of the ancestral species at the root. This is about a 4-fold decrease compared to the number of 25 species, resulting in a confidence interval over 2 times wider than otherwise expected from $n = 25$ i.i.d. sampling units. The analysis of a larger tree with 49 species [Garland et al. (1993)] shows an 8-fold decrease with $n_e = 6.11$. Section 4 shows this is a general phenomenon: increasing the sample size $n$ cannot push the effective sample size $n_e$ associated with the estimation of ancestral states beyond some upper bound. More specifically, Section 4 shows that $n_e \leq kT/t$, where $k$ is the number of edges stemming from the root, $t$ is the length of the shortest of these edges and $T$ is the distance from the root to the tips (or its average value). To account for autocorrelation, Paradis and Claude (2002) introduced a degree of freedom $\mathrm{df}_P = L/T$, where $L$ is the sum of all branch lengths. Interestingly, $n_e$ is necessarily smaller than $\mathrm{df}_P$ when all tips of the tree are at equal distance $T$ from the root (see Appendix A).

Unexpectedly large confidence intervals are already part of biologists' experience [Schluter et al. (1997)]. As Cunningham, Omland and Oakley (1998) put it, likelihood methods have "revealed a surprising amount of uncertainty in ancestral reconstructions" to the point that authors may be tempted to prefer methods that do not report confidence intervals [McArdle and Rodrigo (1994)] or to ignore autocorrelation due to shared ancestry [Martins (2000)]. Still, reconstructing ancestral states or detecting unusual shifts between two ancestors are very frequent goals. For example, Hansen (1997) hypothesized a shift in tooth size to have occurred along the ancient lineage separating browsing horses and grazing horses. Recent micro-array data from gene families have inferred ancestral expression patterns, as well as shifts that possibly occurred after genes were duplicated [Gu (2004)]. Guo et al. (2007) have estimated shifts in brain growth along the human lineage and along the lineage ancestral to human/chimp. Sections 3 and 4 show that under the BM model ancestral reconstructions and shift estimates are not consistent, but are instead convergent toward a random limit. This is illustrated by small effective sample sizes associated with shift estimators. Among the 25 plant species sampled by Davis et al. (2007), 3 parasitic *Rafflesiaceae* species have gigantic flowers (in bold in Figure 3). Under a BM model with a shift on the *Rafflesiaceae* lineage, the effective sample sizes for the root's state ($n_e = 3.98$) and for the shift ($n_e = 2.72$) are obtained from the *Rafflesiaceae* subtree and the remaining subtree. These low effective sample sizes suggest that only large shifts with high power can be detected.

The potential lack of power calls for optimal sampling designs. Trees are typically built from abundant and relatively cheap molecular sequence data. More and more often, a tree comprising many tips is available, while traits of interest cannot be collected from all tips on the tree. A choice has to be made on which tips should be kept for further data collection. Until recently, investigators did not have the tree at hand to make this choice, but now most investigators do. Therefore, optimal sampling design should use information from the tree. Figure 3 shows the effective sample size $n_e$ associated with the root's state in the simple BM model. First, sub-samples were formed by randomly selecting tips and $n_e$ was calculated for each sub-sample. Since there can be a huge number of combinations of tips, 1000 random sub-samples of size $k$ were generated for each $k$. Median and 95% confidence intervals for $n_e$ values are indicated by vertical bars in Figure 3. Second, the sub-samples of a size $k$ that maximize the effective sample size $n_e$ were obtained using step-wise backward and forward searches. Both search strategies agreed on the same maximal $n_e$ values, which are indicated with dots in Figure 3. From both trees, only 15 tips suffice to obtain a near maximum effective sample size, provided that the selected tips are well chosen, not randomly. The proposed selection of tips maximizes $n_e$ and is based on the phylogeny only, prior to data collection. In view of the bound for $n_e$ mentioned above, the selected tips will tend to retain the $k$ edges stemming from the root and to minimize the length of these edges by retaining as many of the early branching lineages as possible.

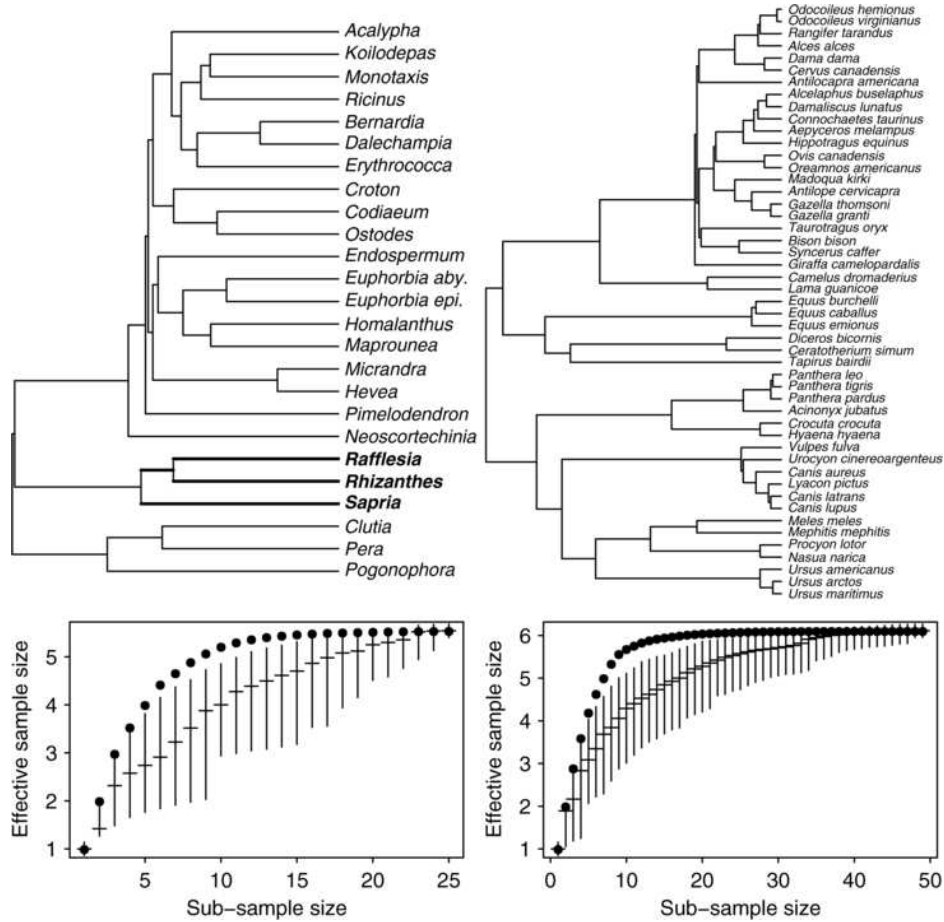

Fig. 3. *Phylogenetic trees from Davis et al. (2007) with 25 plant species, $n_e = 5.54$ (left) and from Garland et al. (1993) with 49 mammal species, $n_e = 6.11$ (right). Bottom: effective sample size $n_e$ for sub-samples of a given size. Vertical bars indicate 95% confidence interval and median $n_e$ values when tips are selected at random from the plant tree (left) and mammal tree (right). Dots indicate optimal $n_e$ values.*

For the purpose of model selection, BIC is widely used [Schwarz (1978), Kass and Raftery (1995), Butler and King (2004)] and is usually defined as $-2 \ln L(\hat{\beta}, \hat{\sigma}) + p \log(n)$, where $L(\hat{\beta}, \hat{\sigma})$ is the maximized likelihood of the model, $p$ the number of parameters and $n$ the number of observations. Each parameter in the model is thus penalized by a $\log(n)$ term. Section 6 shows that this formula does not provide an approximation to the model posterior probability. Instead, the penalty associated with the intercept and with a shift should be bounded, and $\log(1 + n_e)$ is an appropriate penalty to be used for each inconsistent parameter. On the plant tree, the intercept (ancestral value) should therefore be penalized by $\log(1 + 5.54)$ in the simple BM model. In the BM model that includes a shift along the parasitic plant lineage, the intercept should be penalized by $\ln(1 + 3.98)$ and the shift by $\ln(1 + 2.72)$. These penalties are AIC-like (bounded) for high-variance parameters.

## 3. Convergence of estimators.

This section proves the convergence of $\hat{\beta} = \hat{\beta}^{(n)}$ as the sample size $n$ increases. The assumption of a fixed root implies that the covariance matrix $\mathbf{V}_{\text{tree}} = \mathbf{V}_n$ (indexed by the sample size) is a submatrix of $\mathbf{V}_{n+1}$.

THEOREM 1.   *Consider the linear model $Y_i = \mathbf{X}_i \beta + \varepsilon_i$ with*

$$\varepsilon^{(n)} = (\varepsilon_1, \ldots, \varepsilon_n)^t \sim \mathcal{N}(0, \sigma^2 \mathbf{V}_n)$$

*and where predictors $\mathbf{X}$ may be either fixed or random. Assume the design matrix $\mathbf{X}^{(n)}$ (with $\mathbf{X}_i$ for $i$th row) is of full rank provided $n$ is large enough. Then the estimator $\hat{\beta}_n = (\mathbf{X}^{(n)t} \mathbf{V}_n^{-1} \mathbf{X}^{(n)})^{-1} \mathbf{X}^{(n)t} \mathbf{V}_n^{-1} \mathbf{Y}^{(n)}$ is convergent almost surely and in $L^2$. Component $\hat{\beta}_{n,j}$ converges to the true value $\beta_j$ if and only if its asymptotic variance is zero. Otherwise, it converges to a random variable $\hat{\beta}_j^*$, which depends on the tree and the actual data.*

Note that no assumption is made on the covariance structure $\mathbf{V}_n$, except that it is a submatrix of $\mathbf{V}_{n+1}$. Therefore, Theorem 1 holds regardless of how the sequence $\mathbf{V}_n$ is selected. For instance, it holds for the OU model, whose covariance matrix has components $V_{ij} = e^{-\alpha d_{ij}}$ or $V_{ij} = (1 - e^{-2\alpha t_{ij}})e^{-\alpha d_{ij}}$ (depending whether the ancestral state is conditioned upon or integrated out), where $d_{ij}$ is the tree distance between tips $i$ and $j$, and $\alpha$ is the known selection strength.

Theorem 1 can be viewed as a strong law of large numbers: in the absence of covariates and in the i.i.d. case $\hat{\beta}_n$ is just the sample mean. Here, in the absence of covariates $\hat{\beta}_n$ is a weighted average of the observed values, estimating the ancestral state at the root of the tree. Sampling units close to the root could be provided by fossil species or by early viral samples when sampling spans several years. Such units, close to the root, weigh more in $\hat{\beta}_n$ than units further away from the root. Theorem 1 gives a law of large number for this weighted average. However, we will see in Section 4 that the limit is random: $\hat{\beta}_n$ is inconsistent.

PROOF OF THEOREM 1.    The process $\varepsilon = (\varepsilon_1, \varepsilon_2, \ldots)$ is well defined on a probability space $\Omega$ because the covariance matrix $\mathbf{V}_n$ is a submatrix of $\mathbf{V}_{n+1}$. Derivations below are made conditional on the predictors $\mathbf{X}$. In a Bayesian-like approach, the probability space is expanded to $\widetilde{\Omega} = \mathbb{R}^k \times \Omega$ by considering $\beta \in \mathbb{R}^k$ as a random variable, independent of errors $\varepsilon$. Assume a priori that $\beta$ is normally distributed with mean 0 and covariance matrix $\sigma^2 \mathbf{I}_k$, $\mathbf{I}_k$ being the identity matrix of size $k$. Let $\mathcal{F}_n$ be the filtration generated by $Y_1, \ldots, Y_n$. Since $\beta, Y_1, Y_2, \ldots$ is a Gaussian process, the conditional expectation $\mathbb{E}(\beta | \mathcal{F}_n)$ is a linear combination of $Y_1, \ldots, Y_n$ up to a constant:

$$\mathbb{E}(\beta | \mathcal{F}_n) = \mathbf{a}_n + \mathbf{M}_n \mathbf{Y}^{(n)}.$$

The almost sure converge of $\hat{\beta}_n$ will follow from the almost sure convergence of the martingale $\mathbb{E}(\beta | \mathcal{F}_n)$ and from identifying $\mathbf{M}_n \mathbf{Y}^{(n)}$ with a linear transformation of $\hat{\beta}_n$. The vector $\mathbf{a}_n$ and matrix $\mathbf{M}_n$ are such that $\mathbb{E}(\beta | \mathcal{F}_n)$ is the projection of $\beta$ on $\mathcal{F}_n$ in $L^2(\widetilde{\Omega})$, that is, these coefficients are such that

$$\operatorname{trace}(\mathbb{E}(\beta - \mathbf{a}_n - \mathbf{M}_n \mathbf{Y}^{(n)})(\beta - \mathbf{a}_n - \mathbf{M}_n \mathbf{Y}^{(n)})^t)$$

is minimum. Since $Y_i = \mathbf{X}_i \beta + \varepsilon_i$, $\beta$ is centered and independent of $\varepsilon$, we get that $\mathbf{a}_n = 0$ and the quantity to be minimized is

$$\operatorname{tr}((\mathbf{I}_k - \mathbf{M}_n \mathbf{X}^{(n)}) \operatorname{var}(\beta)(\mathbf{I}_k - \mathbf{M}_n \mathbf{X}^{(n)})^t) + \operatorname{tr}(\mathbf{M}_n \operatorname{var}(\epsilon^{(n)}) \mathbf{M}_n^t).$$

The matrix $\mathbf{M}_n$ appears in the first term through $\mathbf{M}_n \mathbf{X}^{(n)}$, so we can minimize $\sigma^2 \operatorname{tr}(\mathbf{M}_n \mathbf{V}_n \mathbf{M}_n^t)$ under the constraint that $\mathbf{B} = \mathbf{M}_n \mathbf{X}^{(n)}$ is fixed. Using Lagrange multipliers, we get $\mathbf{M}_n \mathbf{V}_n = \Lambda \mathbf{X}^{(n)t}$ subject to $\mathbf{M}_n \mathbf{X}^{(n)} = \mathbf{B}$. Assuming $\mathbf{X}^{(n)t} \mathbf{V}_n^{-1} \mathbf{X}^{(n)}$ is invertible, it follows $\Lambda = \mathbf{B}(\mathbf{X}^{(n)t} \mathbf{V}_n^{-1} \mathbf{X}^{(n)})^{-1}$ and $\mathbf{M}_n \mathbf{Y}^{(n)} = \mathbf{B} \hat{\beta}^{(n)}$. The minimum attained is then $\sigma^2 \operatorname{tr}(\mathbf{B}(\mathbf{X}^{(n)t} \mathbf{V}_n^{-1} \mathbf{X}^{(n)})^{-1} \mathbf{B}^t)$. This is necessarily smaller than $\sigma^2 \operatorname{tr}(\mathbf{M} \mathbf{V}_n \mathbf{M}^t)$ when $\mathbf{M}$ is formed by $\mathbf{M}_{n-1}$ and an additional column of zeros. So for any $\mathbf{B}$, the trace of $\mathbf{B}(\mathbf{X}^{(n)t} \mathbf{V}_n^{-1} \mathbf{X}^{(n)})^{-1} \mathbf{B}^t$ is a decreasing sequence. Since it is also nonnegative, it is convergent and so is $(\mathbf{X}^{(n)t}\mathbf{V}_n^{-1}\mathbf{X}^{(n)})^{-1}$. Now the quadratic expression

$$\operatorname{tr}((\mathbf{I}_k - \mathbf{B})(\mathbf{I}_k - \mathbf{B})^t) + \operatorname{tr}(\mathbf{B}(\mathbf{X}^{(n)t}\mathbf{V}_n^{-1}\mathbf{X}^{(n)})^{-1}\mathbf{B}^t)$$

is minimized if $\mathbf{B}$ satisfies $\mathbf{B}(\mathbf{I}_k + (\mathbf{X}^{(n)t}\mathbf{V}_n^{-1}\mathbf{X}^{(n)})^{-1}) = \mathbf{I}_k$. Note the symmetric definite positive matrix $\mathbf{I}_k + (\mathbf{X}^{(n)t}\mathbf{V}_n^{-1}\mathbf{X}^{(n)})^{-1}$ was shown above to be decreasing with $n$. In summary, $\mathbb{E}(\beta|\mathcal{F}_n) = (\mathbf{I}_k + (\mathbf{X}^{(n)t}\mathbf{V}_n^{-1}\mathbf{X}^{(n)})^{-1})^{-1}\hat{\beta}^{(n)}$. This martingale is bounded in $L^2(\widetilde{\Omega})$ so it converges almost surely and in $L^2(\widetilde{\Omega})$ to $\mathbb{E}(\beta|\mathcal{F}_\infty)$. Finally, $\hat{\beta}^{(n)} - \beta = (\mathbf{I}_k + (\mathbf{X}^{(n)t}\mathbf{V}_n^{-1}\mathbf{X}^{(n)})^{-1})\mathbb{E}(\beta|\mathcal{F}_n) - \beta$ is also convergent almost surely and in $L^2(\widetilde{\Omega})$. But $\hat{\beta}_n - \beta$ is a function of $\omega$ in the original probability space $\Omega$, independent of $\beta$. Therefore, for any $\beta$, $\hat{\beta}^{(n)}$ converges almost surely and in $L^2(\Omega)$. Since $\varepsilon$ is a Gaussian process, the limit of $\hat{\beta}^{(n)}$ is normally distributed with covariance matrix the limit of $(\mathbf{X}^{(n)t}\mathbf{V}_n^{-1}\mathbf{X}^{(n)})^{-1}$. It follows that $\hat{\beta}_k^{(n)}$, which is unbiased, converges to the true $\beta_k$ if and only if the $k$th diagonal element of $(\mathbf{X}^{(n)t}\mathbf{V}_n^{-1}\mathbf{X}^{(n)})^{-1}$ goes to 0. $\quad\square$

**4. Consistency of estimators.** In this section I prove bounds on the variance of various effects $\hat{\beta}_i$. From Theorem 1 we know that $\hat{\beta}_i$ is strongly consistent if and only if its variance goes to zero.

4.1. *Intercept.* Assume here that the first column of $\mathbf{X}$ is the column $\mathbf{1}$ of ones, and the first component of $\beta$, the intercept, is denoted $\beta_0$.

PROPOSITION 2. *Let $k$ be the number of daughters of the root node, and let $t$ be the length of the shortest branch stemming from the root. Then* $\operatorname{var}(\hat{\beta}_0) \geq \sigma^2 t/k$. *In particular, when the tree is binary we have* $\operatorname{var}(\hat{\beta}_0) \geq \sigma^2 t/2$.

The following inconsistency result follows directly.

COROLLARY 3. *If there is a lower bound $t > 0$ on the length of branches stemming from the root, and an upper bound on the number of branches stemming from the root, then $\hat{\beta}_0$ is not a consistent estimator of the intercept, even though it is unbiased and convergent.*

The conditions above are very natural in most biological settings, since most ancient lineages have gone extinct. The lower bound may be pushed down if abundant fossil data is available or if there has been adaptive radiation with a burst of speciation events at the root of the tree.

Proof of Proposition 2. Assuming the linear model is correct, the variance of $\hat{\beta}$ is given by $\text{var}(\hat{\beta}) = \sigma^2(\mathbf{X}^t\mathbf{V}^{-1}\mathbf{X})^{-1}$, where the first column of $\mathbf{X}$ is the vector $\mathbf{1}$ of ones, so that the variance of the intercept estimator is just the first diagonal element of $\sigma^2(\mathbf{X}^t\mathbf{V}^{-1}\mathbf{X})^{-1}$. But $(\mathbf{X}^t\mathbf{V}^{-1}\mathbf{X})_{ii}^{-1} \geq (\mathbf{X}_i{}^t\mathbf{V}^{-1}\mathbf{X}_i)^{-1}$ for any index $i$ [Rao (1973), 5a.3, page 327], so the proof can be reduced to the simplest case with no covariates: $Y_i = \beta_0 + \varepsilon_i$. The basic idea is that the information provided by all the tips on the ancestral state $\beta_0$ is no more than the information provided just by the $k$ direct descendants of the root. Let us consider $Z_1, \ldots, Z_k$ to be the character states at the $k$ branches stemming from the root after a time $t$ of evolution (Figure 4, left).

These states are not observed, but the observed values $Y_1, \ldots, Y_n$ have evolved from $Z_1, \ldots, Z_k$. Now I claim that the variance of $\hat{\beta}_0$ obtained from the $Y$ values is no smaller than the variance of $\hat{\beta}_0^{(z)}$ obtained from the $Z$ values. Since the $Z$ values are i.i.d. Gaussian with mean $\beta_0$ and variance $\sigma^2 t$, $\hat{\beta}_0^{(z)}$ has variance $\sigma^2 t/k$. To prove the claim, consider $\beta_0 \sim \mathcal{N}(0, \sigma^2)$ independent of $\epsilon$. Then $\mathbb{E}(\beta_0|Y_1, \ldots, Y_n, Z_1, \ldots, Z_k) = \mathbb{E}(\beta_0|Z_1, \ldots, Z_k)$ so that $\text{var}(\mathbb{E}(\beta_0|Y_1, \ldots, Y_n)) \leq \text{var}(\mathbb{E}(\beta_0|Z_1, \ldots, Z_k))$. The proof of Theorem 1 shows that $\mathbb{E}(\beta_0|Y_1, \ldots, Y_n) = \hat{\beta}_0/(1 + t_y)$ where $t_y = (\mathbf{1}^t\mathbf{V}^{-1}\mathbf{1})^{-1}$ so, similarly, $\mathbb{E}(\beta_0|Z_1, \ldots, Z_k) = \hat{\beta}_0^{(z)}/(1 + t_z)$ where $t_z = t/k$. Since $\beta_0$ and $\hat{\beta}_0 - \beta_0$ are independent, the variance of $\mathbb{E}(\beta_0|Y_1, \ldots, Y_n)$ is $(\sigma^2 + t_y\sigma^2)/(1 + t_y)^2 = \sigma^2/(1 + t_y)$. The variance of $\mathbb{E}(\beta_0|Z_1, \ldots, Z_k)$ is obtained similarly and we get $1/(1 + t_y) \leq 1/(1 + t_z)$, that is, $t_y \geq t_z$ and $\text{var}(\hat{\beta}_0) = \sigma^2(\mathbf{1}^t\mathbf{V}^{-1}\mathbf{1})^{-1} \geq \sigma^2 t/k$. □

4.2. *Lineage effect.* This section considers a predictor $\mathbf{X}_1$ that defines a subtree, that is, $X_{1i} = 1$ if tip $i$ belongs to the subtree and 0 otherwise. This is reduced to a 2-sample comparison problem. The typical "treatment" effect corresponds here to a "lineage" effect, the lineage being the branch subtending the subtree of interest. If a shift occurred along that lineage, tips in the subtree will tend to have, say, high trait values relative to the other tips. However, the BM model does predict a change, on any branch in the tree. So the question is whether the actual shift on the lineage of interest is compatible with a BM change, or whether it is too large to be solely explained by Brownian motion. Alternatively, one might just estimate the actual change.

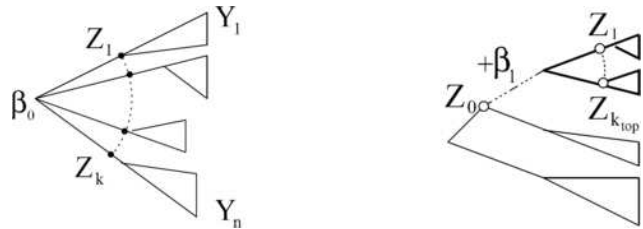

Fig. 4. *Left: Observed states are $Y_1, \ldots, Y_n$, while $Z_1, \ldots, Z_k$ are the unobserved states along the $k$ edges branching from the root, after time $t$ of evolution.* $\mathbf{Y}$ *provides less information on $\beta_0$ than* $\mathbf{Z}$. *Right: $Z_0, Z_1, \ldots, Z_{k_{top}}$ are unobserved states providing more information on the lineage effect $\beta_1$ than the observed $Y$ values.*

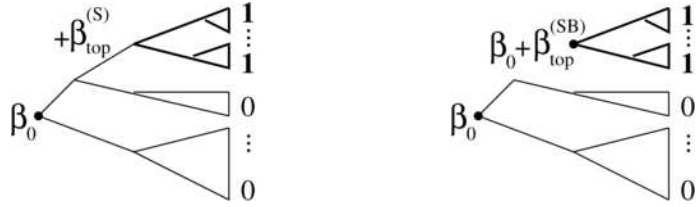

Fig. 5. *Model $M_0$ (left) and $M_1$ (right) with a lineage effect. $\mathbf{X}_1$ is the indicator of a subtree. Model $M_1$ conditions on the state at the subtree's root, modifying the dependence structure.*

This consideration leads to two models. In the first model, a shift $\beta_1 = \beta_{\text{top}}^{(S)}$ is added to the Brownian motion change along the branch of interest, so that $\beta_{\text{top}}^{(S)}$ represents the character displacement *not* due to BM noise. In the second model, $\beta_1 = \beta_{\text{top}}^{(SB)}$ is the actual change, which is the sum of the Brownian motion noise and any extra shift. Observations are then conditioned on the actual ancestral states at the root and the subtree's root (Figure 5). By the Markov property, observations from the two subtrees are conditionally independent of each other. In the second model then, the covariance matrix is modified. The models are written

$$\mathbf{Y} = \mathbf{1}\beta_0 + \mathbf{X}_1\beta_1 + \cdots + \mathbf{X}_k\beta_k + \varepsilon$$

with $\beta_1 = \beta_{\text{top}}^{(S)}$ and $\varepsilon \sim \mathcal{N}(0, \sigma^2 \mathbf{V}_{\text{tree}})$ in the first model, while $\beta_1 = \beta_{\text{top}}^{(SB)}$ and $\varepsilon \sim \mathcal{N}(0, \sigma^2 \operatorname{diag}(\mathbf{V}_{\text{top}}, \mathbf{V}_{\text{bot}}))$ in the second model, where $\mathbf{V}_{\text{top}}$ and $\mathbf{V}_{\text{bot}}$ are the covariance matrices associated with the top and bottom subtrees obtained by removing the branch subtending the group of interest (Figure 5).

PROPOSITION 4. *Let $k_{\text{top}}$ be the number of branches stemming from the subtree of interest, $t_{\text{top}}$ the length of the shortest branch stemming from the root of this subtree, and $t_1$ the length of the branch subtending the subtree. Then*

$$\operatorname{var}(\hat{\beta}_{\text{top}}^{(S)}) \geq \sigma^2(t_1 + t_{\text{top}}/k_{\text{top}}) \quad \text{and} \quad \operatorname{var}(\hat{\beta}_{\text{top}}^{(SB)}) \geq \sigma^2 t_{\text{top}}/k_{\text{top}}.$$

*Therefore, if $t_{\text{top}}/k_{\text{top}}$ remains bounded from below when the sample size increases, both estimators $\hat{\beta}_{\text{top}}^{(S)}$ (pure shift) and $\hat{\beta}_{\text{top}}^{(SB)}$ (actual change) are inconsistent.*

From a practical point of view, unless fossil data is available or there was a radiation (burst of speciation events) at both ends of the lineage, shift estimators are not consistent. Increasing the sample size might not help detect a shift as much as one would typically expect.

Note that the pure shift $\beta_{\text{top}}^{(S)}$ is confounded with the Brownian noise, so it is no wonder that this quantity is not identifiable as soon as $t_1 > 0$. The advantage of the first model is that the BM with no additional shift is nested within it.

PROOF OF PROPOSITION 4. In both models $\text{var}(\hat{\beta}_1)$ is the second diagonal element of $\sigma^2(\mathbf{X}^t\mathbf{V}^{-1}\mathbf{X})^{-1}$ which is bounded below by $\sigma^2(\mathbf{X}_1^t\mathbf{V}^{-1}\mathbf{X}_1)^{-1}$, so that we need just prove the result in the simplest model $\mathbf{Y} = \beta_1\mathbf{X}_1 + \varepsilon$. Similarly to Proposition 2, define $Z_1, \ldots, Z_{k_{\text{top}}}$ as the character states at the $k_{\text{top}}$ direct descendants of the subtree's root after a time $t_{\text{top}}$ of evolution. Also, let $Z_0$ be the state of node just parent to the subtree's root (see Figure 4, right). Like in Proposition 2, it is easy to see that the variance of $\hat{\beta}_1$ given the $\mathbf{Y}$ is larger than the variance of $\hat{\beta}_1$ given the $Z_0, Z_1, \ldots, Z_{k_{\text{top}}}$. In the second model, $\beta_1 = \beta_{\text{top}}^{(SB)}$ is the actual state at the subtree's root, so $Z_1, \ldots, Z_{k_{\text{top}}}$ are i.i.d. Gaussian centered at $\beta_{\text{top}}^{(SB)}$ with variance $\sigma^2 t_{\text{top}}$ and the result follows easily. In the first model, the state at the subtree's root is the sum of $Z_0$, $\beta_{\text{top}}^{(S)}$ and the BM noise along the lineage, so $\hat{\beta}_{\text{top}}^{(S)} = (Z_1 + \cdots + Z_{k_{\text{top}}})/k_{\text{top}} - Z_0$. This estimate is the sum of $\beta_{\text{top}}^{(S)}$, the BM noise and the sampling error about the subtree's root. The result follows because the BM noise and sampling error are independent with variance $\sigma^2 t_1$ and $\sigma^2 t_{\text{top}}/k_{\text{top}}$ respectively. $\square$

4.3. *Variance component.* In contrast to the intercept and lineage effects, inference on the rate $\sigma^2$ of variance accumulation is straightforward. An unbiased estimate of $\sigma^2$ is

$$\hat{\sigma}^2 = \text{RSS}/(n-k) = (\widehat{\mathbf{Y}} - \mathbf{Y})^t\mathbf{V}_{\text{tree}}^{-1}(\widehat{\mathbf{Y}} - \mathbf{Y})/(n-k),$$

where $\widehat{\mathbf{Y}} = \mathbf{X}\hat{\beta}$ are predicted values and $n$ is the number of tips. The classical independence of $\hat{\sigma}^2$ and $\hat{\beta}$ still holds for any tree, and $(n-k)\hat{\sigma}^2/\sigma^2$ follows a $\chi_{n-k}^2$ distribution, $k$ being the rank of $\mathbf{X}$. In particular, $\hat{\sigma}^2$ is unbiased and converges to $\sigma^2$ almost surely as the sample size increases, as shown in Appendix B. Although not surprising, this behavior contrasts with the inconsistency of the intercept and lineage effect estimators. We keep in mind, however, that the convergence of $\hat{\sigma}^2$ may not be robust to a violation of the normality assumption or to a misspecification of the dependence structure, either from a inadequate model (BM versus OU) or from an error in the tree.

4.4. *Random covariate effects.*  In this section $\mathbf{X}$ denotes the matrix of random covariates, excluding the vector of ones or any subtree indicator. In most cases it is reasonable to assume that random covariates also follow a Brownian motion on the tree. Covariates may be correlated, accumulating covariance $t\boldsymbol{\Sigma}$ on any single edge of length $t$. Then covariates $j$ and $k$ have covariance $\Sigma_{jk}\mathbf{V}_{\text{tree}}$. With a slight abuse of notation (considering $\mathbf{X}$ as a single large vector), $\text{var}(\mathbf{X}) = \boldsymbol{\Sigma} \otimes \mathbf{V}_{\text{tree}}$.

PROPOSITION 5.  *Consider $\mathbf{Y} = \mathbf{1}\beta_0 + \mathbf{X}\beta_1 + \varepsilon$ with $\varepsilon \sim \mathcal{N}(0, \sigma^2 \mathbf{V}_{\text{tree}})$. Assume $\mathbf{X}$ follows a Brownian evolution on the tree with nondegenerate covariance $\boldsymbol{\Sigma}$: $\mathbf{X} \sim \mathcal{N}(\mu_X, \boldsymbol{\Sigma} \otimes \mathbf{V}_{\text{tree}})$. Then $\text{var}(\hat{\beta}_1) \sim \sigma^2 \boldsymbol{\Sigma}^{-1}/n$ asymptotically. In particular, $\hat{\beta}_1$ estimates $\beta_1$ consistently by Theorem 1. Random covariate effects are consistently and efficiently estimated, even though the intercept is not.*

PROOF.  We may write $\mathbf{V}^{-1} = \mathbf{R}^t\mathbf{R}$ using the Cholesky decomposition for example. Since $\mathbf{R1} \neq 0$, we may find an orthogonal matrix $\mathbf{O}$ such that $\mathbf{OR1} = (a, 0, \ldots, 0)^t$ for some $a$, so without loss of generality, we may assume that $\mathbf{R1} = (a, 0, \ldots, 0)^t$. The model now becomes $\mathbf{RY} = \mathbf{R1}\beta_0 + \mathbf{RX}\beta_1 + \mathbf{R}\varepsilon$, where errors $\mathbf{R}\varepsilon$ are now i.i.d. Let $\widetilde{\mathbf{X}}_0$ be the first row of $\mathbf{RX}$ and let $\widetilde{\mathbf{X}}_1$ be the matrix made of all but the first row of $\mathbf{RX}$. Similarly, let $(\tilde{y}_0, \widetilde{\mathbf{Y}}_1^t)^t = \mathbf{RY}$ and $(\tilde{\varepsilon}_0, \tilde{\varepsilon}_1^t)^t = \mathbf{R}\varepsilon$. The model becomes $\widetilde{\mathbf{Y}}_1 = \widetilde{\mathbf{X}}_1\beta_1 + \tilde{\varepsilon}_1$, $\tilde{y}_0 = a\beta_0 + \widetilde{\mathbf{X}}_0\beta_1 + \tilde{\varepsilon}_0$ with least square solution $\hat{\beta}_1 = (\widetilde{\mathbf{X}}_1^t\widetilde{\mathbf{X}}_1)^{-1}\widetilde{\mathbf{X}}_1^t\widetilde{\mathbf{Y}}_1 = \beta_1 + (\widetilde{\mathbf{X}}_1^t\widetilde{\mathbf{X}}_1)^{-1}\widetilde{\mathbf{X}}_1^t\tilde{\varepsilon}_1$ and $\hat{\beta}_0 = (\tilde{y}_0 - \widetilde{\mathbf{X}}_0\hat{\beta}_1)/a$. The variance of $\hat{\beta}_1$ conditional on $X$ is then $\sigma^2(\widetilde{\mathbf{X}}_1^t\widetilde{\mathbf{X}}_1)^{-1}$. Using the condition on $\mathbf{R1}$, the rows of $\widetilde{\mathbf{X}}_1$ are i.i.d. centered Gaussian with variance-covariance $\boldsymbol{\Sigma}$ and $(\widetilde{\mathbf{X}}_1^t\widetilde{\mathbf{X}}_1)^{-1}$ has an inverse Wishart distribution with $n-1$ degrees of freedom [Johnson and Kotz (1972)]. The unconditional variance of $\text{var}(\hat{\beta}_1)$ is then $\sigma^2\mathbb{E}(\widetilde{\mathbf{X}}_1^t\widetilde{\mathbf{X}}_1)^{-1} = \sigma^2\boldsymbol{\Sigma}^{-1}/(n-k-2)$, where $k$ is the number of random covariates, which completes the proof.  □

REMARK.  The result still holds if one or more lineage effects are included and if the model conditions upon the character state at each subtree (second model in Section 4.2). The reason is that data from each subtree are independent, and in each subtree the model has just an intercept in addition to the random covariates.

The behavior of random effect estimators contrasts with the behavior of the intercept or lineage effect estimators. An intuitive explanation might be the following. Each cherry in the tree (pair of adjacent tips) is a pair of siblings. Each pair provides independent evidence on the change of $Y$ and of $X$ between the 2 siblings, even though parental information is unavailable. Even though means of $X$ and $Y$ are poorly known, there is abundant evidence on how they change with each other. Similarly, the method of independent contrasts [Felsenstein (1985)] identifies $n-1$ i.i.d. pair-like changes.

**5. Phase transition for symmetric trees.** The motivation for this section is to determine the behavior of the intercept estimator when branches can be sampled closer and closer to the root. I show that the intercept can be consistently estimated, although the rate of convergence can be much slower than root $n$. The focus is on a special case with symmetric sampling (Figure 6). The tree has $m$ levels of internal nodes with the root at level 1. All nodes at level $i$ share the same distance from the root $t_1 + \cdots + t_{i-1}$ and the same number of descendants $d_i$. In a binary tree all internal nodes have 2 descendants and the sample size is $n = 2^m$. The total tree height is set to 1, that is, $t_1 + \cdots + t_m = 1$.

With these symmetries, the eigenvalues of the covariance matrix $\mathbf{V}_n$ can be completely determined (see Appendix C), smaller eigenvalues being associated with shallower internal nodes (close to the tips) and larger eigenvalues being associated with more basal nodes (close to the root). In particular, the constant vector $\mathbf{1}$ is an eigenvector and $(\mathbf{1}^t \mathbf{V}_n \mathbf{1})^{-1} = t_1/d_1 + \cdots + t_m/(d_1 \ldots d_m)$.

In order to sample branches close to the root, consider replicating the major branches stemming from the root. Specifically, a proportion $q$ of each of these $d_1$ branches is kept as is by the root, and the other proportion $1-q$ is replicated along with its subtending tree (Figure 6), that is, $t_1^{(m)} = q^{m-1}$ and $t_i^{(m)} = (1-q)q^{m-i}$ for $i = 2, \ldots, m$. For simplicity, assume further that groups are replicated $d \geq 2$ times at each step, that is, $d_1 = \cdots = d_m = d$.

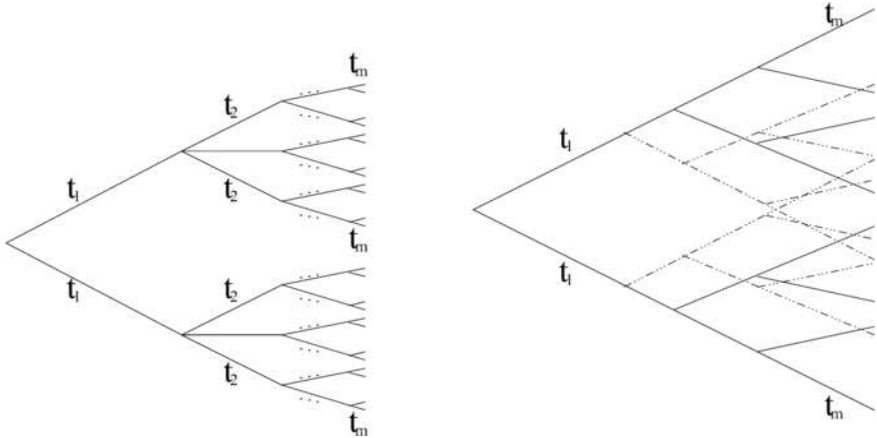

FIG. 6. *Symmetric sampling (left) and replication of major branches close to the root (right).*

The result below shows a law of large numbers and provides the rate of convergence.

PROPOSITION 6. *Consider the model with an intercept and random covariates* $\mathbf{Y} = \mathbf{1}\beta_0 + \mathbf{X}\beta_1 + \varepsilon$ *with* $\varepsilon \sim \mathcal{N}(0, \sigma^2 \mathbf{V}_n)$ *on the symmetric tree described above. Then* $\hat{\beta}_0$ *is consistent. The rate of convergence experiences a phase transition depending on how close to the root new branches are added:* $\mathrm{var}(\hat{\beta}_0)$ *is asymptotically proportional to* $n^{-1}$ *if* $q < 1/d$, $\ln(n)n^{-1}$ *if* $q = 1/d$ *or* $n^\alpha$ *if* $q > 1/d$ *where* $\alpha = \ln(q)/\ln(d)$. *Therefore, the root-n rate of convergence is obtained as in the i.i.d. case if* $q < 1/d$. *Convergence is much slower if* $q > 1/d$.

PROOF. By Theorem 1, the consistency of $\hat{\beta}_0$ follows from its variance going to 0. First consider the model with no covariates. Up to $\sigma^2$, the variance of $\hat{\beta}_0$ is $(\mathbf{1}^t \mathbf{V}_n \mathbf{1})^{-1} = t_1/d_1 + \cdots + t_m/(d_1 \ldots d_m)$, which is $q^{m-1}/d + (1-q)(1-(qd)^{m-1})/(d^m(1-qd))$ if $qd \neq 1$ and $(1+(1-q)(m-1))/d^m$ if $qd = 1$. The result follows easily since $n = d^m$, $m \propto \ln(n)$ and $q^m = n^\alpha$. In the presence of random covariates, it is easy to see that the variance of $\hat{\beta}_0$ is increased by $\mathrm{var}(\hat{\mu}_X(\hat{\beta}_1 - \beta_1))$, where $\hat{\mu}_X = \widetilde{\mathbf{X}}\mathbf{1}/a$ is the row vector of the covariates' estimated ancestral states (using notations from the proof of Proposition 5). By Proposition 5 this increase is $O(n^{-1})$, which completes the proof. □

## 6. Bayesian information criterion. 
The basis for using BIC in model selection is that it provides a good approximation to the marginal model probability given the data and given a prior distribution on the parameters when the sample size is large. The proof of this property uses the i.i.d. assumption quite heavily, and is based on the likelihood being more and more peaked around its maximum value. Here, however, the likelihood does not concentrate around its maximum value since even an infinite sample size may contain little information about some parameters in the model. The following proposition shows that the penalty associated with the intercept or with a lineage effect ought to be bounded, thus smaller than $\log(n)$.

PROPOSITION 7. *Consider* $k$ *random covariates* $\mathbf{X}$ *with Brownian evolution on the tree and nonsingular covariance* $\mathbf{\Sigma}$, *and the linear models*

$$\mathbf{Y} = \beta_0 \mathbf{1} + \mathbf{X}\beta_1 + \varepsilon \qquad \text{with } \varepsilon \sim \mathcal{N}(0, \sigma^2 \mathbf{V}_{\text{tree}}) \qquad (\mathrm{M}_0)$$

$$\mathbf{Y} = \beta_0 \mathbf{1} + \mathbf{X}\beta_1 + \beta_{\text{top}}\mathbf{1}_{\text{top}} + \varepsilon \qquad \text{with } \varepsilon \sim \mathcal{N}(0, \sigma^2 \mathbf{V}_{\text{tree}}), \qquad (\mathrm{M}_1)$$

*where the lineage factor* $\mathbf{1}_{\text{top}}$ *is the indicator of a (top) subtree. Assume a smooth prior distribution* $\pi$ *over the parameters* $\theta = (\beta, \sigma)$ *and a sampling such that* $\mathbf{1}^t \mathbf{V}_n^{-1} \mathbf{1}$ *is bounded, that is, branches are not sampled too close* *from the root. With model $M_1$ assume further that branches are not sampled too close from the lineage of interest, that is, $\mathbf{1}_{\mathrm{top}}^t \mathbf{V}_n^{-1} \mathbf{1}_{\mathrm{top}}$ is bounded. Then for both models, the marginal probability of the data $\mathbb{P}(Y) = \int \mathbb{P}(Y|\theta)\pi(\theta)\,d\theta$ satisfies*

$$-2\log\mathbb{P}(Y) = -2\ln L(\hat{\theta}) + (k+1)\ln(n) + O(1)$$

*as the sample size increases. Therefore, the penalty for the intercept and for a lineage effect is bounded as the sample size increases.*

The poorly estimated parameters are not penalized as severely as the consistently estimated parameters, since they lead to only small or moderate increases in likelihood. Also, the prior information continues to influence the posterior of the data even with a very large sample size. Note that the lineage effect $\beta_{\mathrm{top}}$ may either be the pure shift or the actual change. Model $M_0$ is nested within $M_1$ in the first case only.

In the proof of Proposition 7 (see Appendix D) the $O(1)$ term is shown to be dominated by

$$C = \log\det\hat{\boldsymbol{\Sigma}} - (k+1)\log(2\pi\hat{\sigma}^2) + \log 2 + D,$$

where $D$ depends on the model. In $M_0$

$$(1)\qquad D = -2\log\int_{\beta_0}\exp(-(\beta_0 - \hat{\beta}_0)^2/(2t_0\hat{\sigma}^2))\pi(\beta_0, \hat{\beta}_1, \hat{\sigma})\,d\beta_0,$$

where $t_0 = \lim(\mathbf{1}^t\mathbf{V}_n^{-1}\mathbf{1})^{-1}$. In $M_1$

$$(2)\quad D = -2\log\int_{\beta_0, \beta_{\mathrm{top}}}\exp(-\tilde{\beta}^t\mathbf{W}^{-1}\tilde{\beta}/(2\hat{\sigma}^2))\pi(\beta_0, \beta_{\mathrm{top}}, \hat{\beta}_1, \hat{\sigma})\,d\beta_0\,d\beta_{\mathrm{top}},$$

where $\tilde{\beta}^t = (\beta_0 - \hat{\beta}_0, \beta_{\mathrm{top}} - \hat{\beta}_{\mathrm{top}})$ and the $2\times 2$ symmetric matrix $\mathbf{W}^{-1}$ has diagonal elements $\lim\mathbf{1}^t\mathbf{V}_n^{-1}\mathbf{1} = t_0^{-1}$, $\lim\mathbf{1}_{\mathrm{top}}^t\mathbf{V}_n^{-1}\mathbf{1}_{\mathrm{top}} < \infty$ and off-diagonal element $\lim\mathbf{1}^t\mathbf{V}_n^{-1}\mathbf{1}_{\mathrm{top}}$, which does exist.

In the rest of the section I assume that all tips are at the same distance $T$ from the root. This condition is realized when branch lengths are chronological times and tips are sampled simultaneously. Under BM, $Y_1, \ldots, Y_n$ have common variance $\sigma^2 T$. The ancestral state at the root is estimated with asymptotic variance $\sigma^2/\lim_n \mathbf{1}^t\mathbf{V}_n^{-1}\mathbf{1}$, while the same precision would be obtained with a sample of $n_e$ independent variables where

$$n_e = T\lim_n \mathbf{1}^t\mathbf{V}_n^{-1}\mathbf{1}.$$

Therefore, I call this quantity the *effective sample size* associated with the intercept.

The next proposition provides more accuracy for the penalty term in case the prior has a specific, reasonable form. In some settings, it has been shown that the error term in the BIC approximation is actually better than $O(1)$. Kass and Wasserman (1995) show this error term is only $O(n^{-1/2})$ if the prior carries the same amount of information as a single observation would, as well as in the context of comparing nested models with an alternative hypothesis close to the null. I follow Kass and Wasserman (1996) and consider a "reference prior" that contains little information, like a single observation would [see also Raftery (1995, 1996), Wasserman (2000)]. In an empirical Bayes way, assume the prior is Gaussian centered at $\hat{\theta}$. Let $(\beta_1, \sigma)$ have prior variance $\mathbf{J}_n^{-1} = \mathrm{diag}(\hat{\sigma}^2\hat{\boldsymbol{\Sigma}}^{-1}, \hat{\sigma}^2/2)$ and be independent of the other parameter(s) $\beta_0$ and $\beta_{\mathrm{top}}$. Also, let $\beta_0$ have variance $\hat{\sigma}^2 T$ in model $M_0$.

In model $M_1$, assume further that the tree is rooted at the base of the lineage of interest, so that the intercept is the ancestral state at the base of that lineage. This reparametrization has the advantage that $\hat{\beta}_0$ and $\hat{\beta}_0 + \hat{\beta}_{\mathrm{top}}$ are uncorrelated asymptotically. A single observation from outside the subtree of interest (i.e., from the bottom subtree) would be centered at $\beta_0$ with variance $\sigma^2 T$, while a single observation from the top subtree would be centered at $\beta_0 + \beta_{\mathrm{top}}$ with variance $\sigma^2 T_{\mathrm{top}}$. In case $\beta_{\mathrm{top}}$ is the pure shift, then $T_{\mathrm{top}} = T$. If $\beta_{\mathrm{top}}$ is the actual change along the lineage, then $T_{\mathrm{top}}$ is the height of the subtree excluding its subtending branch. Therefore, it is reasonable to assign $(\beta_0, \beta_{\mathrm{top}})$ a prior variance of $\hat{\sigma}^2 \mathbf{W}_\pi$ with

$$\mathbf{W}_\pi = \begin{pmatrix} T & -T \\ -T & T + T_{\mathrm{top}} \end{pmatrix}.$$

The only tips informing $\beta_0 + \beta_{\mathrm{top}}$ are those in the top subtree and the only units informing $\beta_0$ are those in the bottom subtree. Therefore, the effective sample sizes associated with the intercept and lineage effects are defined as

$$n_{e,\mathrm{bot}} = T \lim_n \mathbf{1}^t \mathbf{V}_{\mathrm{bot}}^{-1} \mathbf{1}, \qquad n_{e,\mathrm{top}} = T_{\mathrm{top}} \lim_n \mathbf{1}^t \mathbf{V}_{\mathrm{top}}^{-1} \mathbf{1},$$

where $\mathbf{V}_{\mathrm{bot}}$ and $\mathbf{V}_{\mathrm{top}}$ are the variance matrices from the bottom and top subtrees.

PROPOSITION 8. *Consider models $M_0$ and $M_1$ and the prior specified above. Then $\mathbb{P}(Y|\mathrm{M}_0) = -2\ln L(\hat{\theta}|\mathrm{M}_0) + (k+1)\ln(n) + \ln(1+n_e) + o(1)$ and $\mathbb{P}(Y|\mathrm{M}_1) = -2\ln L(\hat{\theta}|\mathrm{M}_1) + (k+1)\ln(n) + \ln(1+n_{e,\mathrm{bot}}) + \ln(1+n_{e,\mathrm{top}}) + o(1)$. Therefore, a reasonable penalty for the nonconsistently estimated parameters is the log of their effective sample sizes plus one.*

PROOF.    With model $M_0$, we get from (1)

$$D = -2\log\pi(\hat{\beta}_1, \hat{\sigma}) - 2\log\int \exp\left(-\frac{(\beta_0-\hat{\beta}_0)^2}{2t_0\hat{\sigma}^2} - \frac{(\beta_0-\hat{\beta}_0)^2}{2T\hat{\sigma}^2}\right)\frac{d\beta_0}{\sqrt{2\pi T}\hat{\sigma}}$$

$$= -2\log\pi(\hat{\beta}_1, \hat{\sigma}) + \log(1 + T/t_0).$$

Now $-2 \log \pi(\hat{\beta}_1, \hat{\sigma}) = (k+1) \log(2\pi) - \log \det \mathbf{J}_n$ cancels with the first constant terms to give $C = \log(1 + T/t_0) = \log(1 + n_e)$. With model $M_1$, we get

$$D = -2 \log \pi(\hat{\beta}, \hat{\sigma}) - 2 \log \frac{\det(\mathbf{W}^{-1} + \mathbf{W}_\pi^{-1})^{-1/2}}{\det \mathbf{W}_\pi^{1/2}},$$

so that again $C = \log(\det(\mathbf{W}^{-1} + \mathbf{W}_\pi^{-1}) \det \mathbf{W}_\pi)$. It remains to identify this quantity with $\ln(1 + n_{e,\text{bot}}) + \ln(1 + n_{e,\text{top}})$. It is easy to see that $\det \mathbf{W}_\pi = T T_\text{top}$ and

$$\mathbf{W}_\pi^{-1} = \begin{pmatrix} T^{-1} + T_\text{top}^{-1} & T_\text{top}^{-1} \\ T_\text{top}^{-1} & T_\text{top}^{-1} \end{pmatrix}.$$

Since $\mathbf{V}$ is block diagonal $\text{diag}(\mathbf{V}_\text{top}, \mathbf{V}_\text{bot})$, we have that $\mathbf{1}^t \mathbf{V}^{-1} \mathbf{1}_\text{top} = n_{e,\text{top}}/T_\text{top}$ and $\mathbf{1}^t \mathbf{V}^{-1} \mathbf{1} = \mathbf{1}^t \mathbf{V}_\text{top}^{-1} \mathbf{1} + \mathbf{1}^t \mathbf{V}_\text{bot}^{-1} \mathbf{1} = n_{e,\text{top}}/T_\text{top} + n_{e,\text{bot}}/T$. Therefore, $\mathbf{W}^{-1}$ has diagonal terms $n_{e,\text{top}}/T_\text{top} + n_{e,\text{bot}}/T$ and $n_{e,\text{bot}}/T_\text{top}$ and off-diagonal term $n_{e,\text{bot}}/T_\text{top}$. We get $\det(\mathbf{W}^{-1} + \mathbf{W}_\pi^{-1}) = (n_{e,\text{bot}} + 1)/T_\text{top}(n_{e,\text{bot}} + 1)/T$, which completes the proof. $\square$

*Akaike's information criterion (AIC).* This criterion [Akaike (1974)] is also widely used for model selection. With i.i.d. samples, AIC is an estimate of the Kullback–Leibler divergence between the true distribution of the data and the estimated distribution, up to a constant [Burnham and Anderson (2002)]. Because of the BM assumption, the Kullback–Leibler divergence can be calculated explicitly. Using the Gaussian distribution of the data, the mutual independence of $\hat{\sigma}^2$ and $\hat{\beta}$ and the chi-square distribution of $\hat{\sigma}^2$, the usual derivation of AIC applies. Contrary to BIC, the AIC approximation still holds with tree-structure dependence.

**7. Applications and discussion.** This paper provides a law of large numbers for non i.i.d. sequences, whose dependence is governed by a tree structure. Almost sure convergence is obtained, but the limit may or may not be the expected value. With spatial or temporal data, the correlation decreases rapidly with spatial distance or with time typically (e.g., AR processes) under expanding asymptotics. With a tree structure, the dependence of any 2 *new* observations from 2 given subtrees will have the same correlation with each other as with "older" observations. In spatial statistics, infill asymptotics also harbor a strong, nonvanishing correlation. This dependence implies a bounded effective sample size $n_e$ in most realistic biological settings. However, I showed that this effective sample size pertains to locations parameters only (intercept, lineage effects). Inconsistency has also been described in population genetics. In particular, Tajima's estimator of the level of sequence diversity from a sample of $n$ individuals is not consistent [Tajima (1983)], while asymptotically optimal estimators only converge at rate $\log(n)$ rather than $n$ [Fu and Li (1993)]. The reason is that the genealogical correlation among individuals in the population decreases the available information.

*Sampling design.* Very large genealogies are now available, with hundreds or thousands of tips [Cardillo et al. (2005), Beck et al. (2006)]. It is not uncommon that physiological, morphological or other phenotypic data cannot be measured for all units in the group of interest. For the purpose of estimating an ancestral state, the sampling strategy suggested here maximizes the scaled effective sample size $\mathbf{1}^t \mathbf{V}_n^{-1} \mathbf{1}$ over all subsamples of size $n$, where $n$ is an affordable number of units to subsample. This criterion is a function of the rooted tree topology and its branch lengths. It is very easy to calculate with one tree traversal using Felsenstein's algorithm [Felsenstein (1985)], without inverting $\mathbf{V}_n$. It might be computationally too costly to assess all subsamples of size $n$, but one might heuristically search only among the most star-like subtrees. Backward and forward stepwise search strategies were implemented, either removing or adding tips one at a time.

*Desperate situation?* This paper provides a theoretical explanation for the known difficulty of estimating ancestral states. In terms of detecting non-Brownian shifts, our results imply that the maximum power cannot reach 100%, even with infinite sampling. Instead, what mostly drives the power of shift detection is the effect size: $\beta_1/\sqrt{t}\sigma$ where $\beta_1$ is the shift size and $t$ is the length of the lineage experiencing the shift. The situation is desperate only in cases when the effect size is small. Increased sampling may not provide more power.

*Beyond the Brownian motion model.* The convergence result applies to any dependence matrix. Bounds on the variance of estimates do not apply to the Ornstein–Uhlenbeck model, so it would be interesting to study the consistency of estimates in this model. Indeed, when selection is strong the OU process is attracted to the optimal value and "forgets" the initial value exponentially fast. Several studies have clearly indicated that some ancestral states and lineage-specific optimal values are not estimable [Butler and King (2004), Verdú and Gleiser (2006)], thus bearing on the question of how efficiently these parameters can be estimated. While the OU model is already being used, theoretical questions remain open.

*Broader hierarchical autocorrelation context.* So far linear models were considered in the context of biological data with shared ancestry. However, implications of this work are far reaching and may affect common practices in many fields, because tree structured autocorrelation underlies many experimental designs. For instance, the typical ANOVA can be represented by a forest (with BM evolution), one star tree for each group (Figure 7). If groups have random effects, then a single tree captures this model (Figure 7). It shows visually how the variation decomposes into within and among group variation. ANOVA with several nested effects would be represented by a tree with more hierarchical levels, each node in the tree representing a group. In such random (or mixed) effect models, asymptotic results are known when the number of groups becomes large, while the number of units per group is not necessarily required to grow [Akritas and Arnold (2000), Wang and Akritas (2004), Güven (2006)]. The results presented here pertain to any kind of tree growth, even when group sizes are bounded.

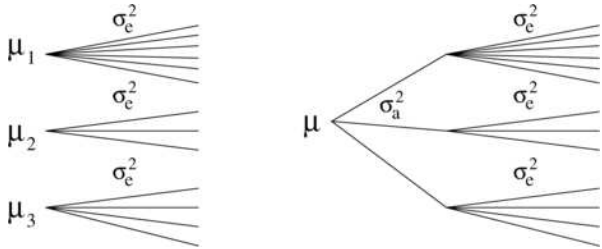

Fig. 7.   *Trees associated with ANOVA models: 3 groups with fixed effects (left) or random effects (right). Variance within and among groups are $\sigma_e^2$ and $\sigma_a^2$ respectively.*

*Model selection.*   Many aspects of the model can be selected for, such as the most important predictors or the appropriate dependence structure. Moreover, there often is some uncertainty in the tree structure or in the model of evolution. Several trees might be obtained from molecular data on several genes, for instance. These trees might have different topologies or just different sets of branch lengths. BIC values from several trees can be combined for model averaging. I showed in this paper that the standard form of BIC is inappropriate. Instead, I propose to adjust the penalty associated to an estimate with its effective sample size. AIC was shown to be still appropriate for approximating the Kullback–Leibler criterion.

*Open questions.*   It was shown that the scaled effective sample size is bounded as long as the number $k$ of edges stemming from the root is bounded and their lengths are above some $t > 0$. The converse is not true in general. Take a star tree with edges of length $n^2$. Then $Y_n \sim \mathcal{N}(\mu, \sigma^2 n^2)$ are independent, and $\mathbf{1}^t \mathbf{V}_n^{-1} \mathbf{1} = \sum n^{-2}$ is bounded. However, if one requires that the tree height is bounded (i.e., tips are distant from the root by no more than a maximum amount), then is it necessary to have $k < \infty$ and $t > 0$ for the effective sample size to be bounded? If not, it would be interesting to know a necessary condition.

## APPENDIX A: UPPER BOUND FOR THE EFFECTIVE SAMPLE SIZE

I prove here the claim made in Section [2] that the effective sample size for the intercept $n_e = T \mathbf{1}^t \mathbf{V}^{-1} \mathbf{1}$ is bounded by $\mathrm{df}_P = L/T$, where $L$ is the tree length (the sum of all branch lengths), in case all tips are at equal distance $T$ from the root. It is easy to see that $\mathbf{V}$ is block diagonal, each block corresponding to one subtree branching from the root. Therefore, $\mathbf{V}^{-1}$ is also block diagonal and, by induction, we only need to show that $n_e \leq L/T$ when the root is adjacent to a single edge. Let $t$ be the length of this edge. When this edge is pruned from the tree, one obtains a subtree of length $L - t$ and whose tips are at distance $T - t$ from the root. Let $\mathbf{V}_{-t}$ be the covariance matrix associated with this subtree. By induction, one may assume that $\mathbf{1}^t \mathbf{V}_{-t} \mathbf{1} \leq (L - t)/(T - t)^2$. Now $\mathbf{V}$ is of the form $t\mathbf{J} + \mathbf{V}_{-t}$, where $\mathbf{J} = \mathbf{1}\mathbf{1}^t$ is a square matrix of ones. It is easy to check that $\mathbf{V}^{-1}\mathbf{1} = \mathbf{V}_{-t}^{-1}\mathbf{1}/(1 + t\mathbf{1}^t \mathbf{V}_{-t}^{-1}\mathbf{1})$ so that $\mathbf{1}^t \mathbf{V}^{-1}\mathbf{1} = ((\mathbf{1}^t\mathbf{V}_{-t}^{-1}\mathbf{1})^{-1} + t)^{-1} \leq ((T-t)^2/(L-t) + t)^{-1}$. By concavity of the inverse function, $((1-\lambda)/a + \lambda/b)^{-1} < (1-\lambda)a + \lambda b$ for all $\lambda$ in $(0, 1)$ and all $a > b > 0$. Combining the two previous inequalities with $\lambda = t/T$, $a = (L-t)/(T-t)$ and $b = 1$ yields $\mathbf{1}^t \mathbf{V}^{-1}\mathbf{1} < L/T^2$ and proves the claim. The equality $n_e = \mathrm{df}_P$ only occurs when the tree is reduced to a single tip, in which case $n_e = 1 = \mathrm{df}_P$.

## APPENDIX B: ALMOST SURE CONVERGENCE OF $\hat{\sigma}$ AND $\hat{\Sigma}$

Convergence of $\hat{\sigma}$ in probability is obtained because $\nu\hat{\sigma}_n^2/\sigma^2$ has a chi-square distribution with degree of freedom $\nu = n - r$, $r$ being the total number of covariates. The exact knowledge of this distribution provides bounds on tail probabilities. Strong convergence follows from the convergence of the series $\sum_n \mathbb{P}(|\hat{\sigma}_n^2 - \sigma^2| > \varepsilon) < \infty$ for all $\varepsilon > 0$, which in turn follows from the application of Chernov's bound and derivation of large deviations [Dembo and Zeitouni (1998)]: $\mathbb{P}(\hat{\sigma}^2 - \sigma^2 > \varepsilon) \leq \exp(-\nu I(\varepsilon))$ and $\mathbb{P}(\hat{\sigma}^2 - \sigma^2 < -\varepsilon) \leq \exp(-\nu I(-\varepsilon))$ where the rate function $I(\varepsilon) = (\varepsilon - \log(1 + \varepsilon))/2$ for all $\varepsilon > -1$ is obtained from the moment generating function of the chi-square distribution.

The covariance matrix of random effects is estimated with $\nu\hat{\Sigma}_n = \widetilde{\mathbf{X}}_1^t \widetilde{\mathbf{X}}_1 = (\mathbf{X} - \hat{\mu}_X)^t \mathbf{V}_n^{-1}(\mathbf{X} - \hat{\mu}_X)$, with $\widetilde{\mathbf{X}}_1$ as in the proof of Proposition [5], which has a Wishart distribution with degree of freedom $\nu = n - 1$ and variance parameter $\mathbf{\Sigma}$. For each vector $\mathbf{c}$ then, $\mathbf{c}^t \nu\hat{\mathbf{\Sigma}}_\mathbf{n}\mathbf{c}$ has a chi-square distribution with variance parameter $\mathbf{c}^t\mathbf{\Sigma}\mathbf{c}$, so that $\mathbf{c}^t\hat{\Sigma}_n\mathbf{c}$ converges almost surely to $\mathbf{c}^t\mathbf{\Sigma}\mathbf{c}$ by the above argument. Using the indicator of the $j$th coordinate $\mathbf{c} = \mathbf{1}_j$, then $\mathbf{c} = \mathbf{1}_i + \mathbf{1}_j$, we obtain the strong convergence of $\hat{\mathbf{\Sigma}}$ to $\mathbf{\Sigma}$.

## APPENDIX C: SYMMETRIC TREES

With the symmetric sampling from Section 5, eigenvalues of $\mathbf{V}_n$ are of the form

$$\lambda_i = n\left(\frac{t_i}{d_1 \ldots d_i} + \cdots + \frac{t_m}{d_1 \ldots d_m}\right)$$

with multiplicity $d_1 \ldots d_{i-1}(d_i - 1)$, the number of nodes at level $i$ if $i \geq 2$. At the root (level 1) the multiplicity is $d_1$. Indeed, it is easy to exhibit the eigenvectors of each $\lambda_i$. Consider $\lambda_1$ for instance. The $d_1$ descendants of the root define $d_1$ groups of tips. If $\mathbf{v}$ is a vector such that $v_j = v_k$ for tips $j$ and $k$ is the same group, then it is easy to see that $\mathbf{V}_n\mathbf{v} = \lambda_1\mathbf{v}$. It shows that $\lambda_1$ is an eigenvalue with multiplicity $d_1$ (at least). Now consider an internal node at level $i$. Its descendants form $d_i$ groups of tips, which we name $G_1, \ldots, G_{d_i}$. Let $\mathbf{v}$ be a vector such that $v_j = 0$ if tip $j$ is not a descendant of the node and $v_j = a_g$ if $j$ is a descendant from group $g$. Then, if $a_1 + \cdots + a_{d_i} = 0$, it is easy to see that $\mathbf{V}_n\mathbf{v}\lambda_i\mathbf{v}$. Since the multiplicities sum to $n$, all eigenvalues and eigenvectors have been identified.

## APPENDIX D: BIC APPROXIMATION

PROOF OF PROPOSITION 7. The prior $\pi$ is assumed to be sufficiently smooth (four times continuously differentiable) and bounded. The same conditions are also required for $\pi_m$ defined by $\pi_m = \sup_{\beta_0} \pi(\beta_1, \sigma | \beta_0)$ in model $M_0$ and $\pi_m = \sup_{\beta_0, \beta_{\text{top}}} \pi(\beta_1, \sigma | \beta_0, \beta_{\text{top}})$ in model $M_1$. The extra assumption on $\pi_m$ is pretty mild; it holds when parameters are independent a priori, for instance.

For model $M_0$ the likelihood can be written

$$-2\log L(\theta) = -2\log L(\hat{\theta}) + n\left(\frac{\hat{\sigma}^2}{\sigma^2} - 1 - \log\frac{\hat{\sigma}^2}{\sigma^2}\right)$$

$$+ ((\beta_1 - \hat{\beta}_1)^t\mathbf{X}^t\mathbf{V}_n^{-1}\mathbf{X}(\beta_1 - \hat{\beta}_1) + \mathbf{1}^t\mathbf{V}_n^{-1}\mathbf{1}(\beta_0 - \hat{\beta}_0)^2$$

$$+ 2(\beta_0 - \hat{\beta}_0)\mathbf{1}^t\mathbf{V}_n^{-1}\mathbf{X}(\beta_1 - \hat{\beta}_1))/\sigma^2.$$

Rearranging terms, we get $-2\log L(\theta) = -2\log L(\hat{\theta}) + 2nh_n(\theta) + a_n(\beta_0 - \hat{\beta}_0)^2/\hat{\sigma}^2$, where $a_n = \mathbf{1}^t\mathbf{V}_n^{-1}\mathbf{1} - \mathbf{1}^t\mathbf{V}_n^{-1}\mathbf{X}(\mathbf{X}^t\mathbf{V}_n^{-1}\mathbf{X})^{-1}\mathbf{X}^t\mathbf{V}_n^{-1}\mathbf{1}$,

$$2h_n(\theta) = \left(\frac{\hat{\sigma}^2}{\sigma^2} - 1 - \log\frac{\hat{\sigma}^2}{\sigma^2}\right) + (\beta_1 - u_1)^t\frac{\mathbf{X}^t\mathbf{V}_n^{-1}\mathbf{X}}{n\sigma^2}(\beta_1 - u_1)$$

$$+ \frac{a_n}{n}(\beta_0 - \hat{\beta}_0)^2\left(\frac{1}{\sigma^2} - \frac{1}{\hat{\sigma}^2}\right)$$

and $u_1 = \hat{\beta}_1 - (\beta_0 - \hat{\beta}_0)(\mathbf{X}^t\mathbf{V}_n^{-1}\mathbf{X})^{-1}\mathbf{X}^t\mathbf{V}_n^{-1}\mathbf{1}$. For any fixed value of $\beta_0$, consider $h_n$ as a function of $\beta_1$ and $\sigma$. Its minimum is attained at $u_1$ and $\hat{\sigma}_1^2 = \hat{\sigma}^2 + a_n(\beta_0 - \hat{\beta}_0)^2/n$. At this point the minimum value is $2h_n(u_1, \hat{\sigma}_1) = \log(1 + a_n(\beta_0 - \hat{\beta}_0)^2/(n\hat{\sigma}^2)) - a_n(\beta_0 - \hat{\beta}_0)^2/(n\hat{\sigma}^2)$ and the second derivative of $h_n$ is $\mathbf{J}_n = \mathrm{diag}(\mathbf{X}^t\mathbf{V}_n^{-1}\mathbf{X}/(n\hat{\sigma}_1^2), 2/\hat{\sigma}_1^2)$. Note that $\hat{\mu}_X = \mathbf{1}^t\mathbf{V}_n^{-1}\mathbf{X}/(\mathbf{1}^t\mathbf{V}_n^{-1}\mathbf{1})$ is the row vector of estimated ancestral states of $\mathbf{X}$, so by Theorem 1, it is convergent. Note also that $\mathbf{X}^t\mathbf{V}_n^{-1}\mathbf{X} = (n-1)\hat{\boldsymbol{\Sigma}} + (\mathbf{1}^t\mathbf{V}_n^{-1}\mathbf{1})\mu_X{}^t\mu_X$. Since $\mathbf{1}^t\mathbf{V}_n^{-1}\mathbf{1}$ is assumed bounded, $\mathbf{X}^t\mathbf{V}_n^{-1}\mathbf{X} = n\hat{\boldsymbol{\Sigma}} + O(1)$ almost surely, and the error term depends on $\mathbf{X}$ only, not on the parameters $\beta$ or $\sigma$. Consequently, $a_n = \mathbf{1}^t\mathbf{V}_n^{-1}\mathbf{1} + O(n^{-1})$ is almost surely bounded and $\hat{\sigma}_1^2 = \hat{\sigma}^2 + O(n^{-1})$. It follows that for any fixed $\beta_0$, $\mathbf{J}_n$ converges almost surely to $\mathrm{diag}(\boldsymbol{\Sigma}/\sigma^2, 2/\sigma^2)$. Therefore, its eigenvalues are almost surely bounded and bounded away from zero, and $h_n$ is Laplace-regular as defined in Kass, Tierney and Kadane (1990). Theorem 1 in Kass, Tierney and Kadane (1990) shows that

$$-2\log \int e^{-nh_n} \pi \, d\beta_1 \, d\sigma$$

$$= 2nh_n(u_1, \hat{\sigma}_1) + (k+1)\log n + \log \det \hat{\boldsymbol{\Sigma}}_1$$

$$- (k+1)\log(2\pi\hat{\sigma}_1^2) + \log 2 - 2\log(\pi(\hat{\beta}_1, \hat{\sigma}|\beta_0) + O(n^{-1}))$$

with $\hat{\boldsymbol{\Sigma}}_1 = \mathbf{X}^t\mathbf{V}_n^{-1}\mathbf{X}/n = \hat{\boldsymbol{\Sigma}} + O(n^{-1})$. Integrating further over $\beta_0$, we get $-2\log \mathbb{P}(Y) = -2\log L(\hat{\theta}) + (k+1)\log n + \log \det \hat{\boldsymbol{\Sigma}}_1 - (k+1)\log(2\pi\hat{\sigma}^2) + \log 2 + D_n$, where

$$D_n = -2\log \int \exp\left(-\frac{n-k-1}{2}\log\left(1 + \frac{a_n(\beta_0 - \hat{\beta}_0)^2}{n\hat{\sigma}^2}\right)\right)$$

$$\times (\pi(\hat{\beta}_1, \hat{\sigma}|\beta_0) + O(n^{-1}))\pi(\beta_0) \, d\beta_0.$$

Heuristically, we see that $a_n$ converges to $t_0^{-1} = \lim \mathbf{1}^t\mathbf{V}_n^{-1}\mathbf{1}$ and for fixed $\beta_0$ the integrand is equivalent to $\exp(-(\beta_0 - \hat{\beta}_0)^2/(2t_0\hat{\sigma}^2))$, so we would conclude that $D_n$ converges to $D = -2\log \int \exp(-(\beta_0 - \hat{\beta}_0)^2/(2t_0\hat{\sigma}^2))\pi(\beta_0, \hat{\beta}_1, \hat{\sigma}) \, d\beta_0$ as given in (1) and, thus,

$$-2\log \mathbb{P}(Y) = -2\log L(\hat{\theta}) + (k+1)\log n + \log \det \hat{\boldsymbol{\Sigma}} - (k+1)\log(2\pi\hat{\sigma}^2)$$

$$+ \log 2 + D + o(1).$$

Formally, we need to check that the $O(n^{-1})$ term in $D_n$ has an $o(1)$ contribution after integration, and that the limit of the integral is the integral of the point-wise limit. The integrand in $D_n$ is the product of

$$f_n(\beta_0) = n^{(k+1)/2} \int \exp(\log L(\theta) - \log L(\hat{\theta}))\pi(\beta_1, \sigma|\beta_0) \, d\beta_1 \, d\sigma$$

and of a quantity that converges almost surely: $(2\det \hat{\boldsymbol{\Sigma}}_1)^{1/2}(2\pi\hat{\sigma}^2)^{-(k+1)/2}$. Maximizing the likelihood and prior in $\beta_0$, we get that $f_n$ is uniformly

bounded in $\beta_0$ by

$$n^{(k+1)/2} \int \exp\left(-\frac{n}{2}\left(\frac{\hat{\sigma}^2}{\sigma^2} - 1 - \log\frac{\hat{\sigma}^2}{\sigma^2} + (\beta_1 - \hat{\beta}_1)^t \hat{\boldsymbol{\Sigma}}_2 (\beta_1 - \hat{\beta}_1)/\sigma^2\right)\right)$$
$$\times \pi_m(\beta_1, \sigma)\, d\beta_1\, d\sigma,$$

where $\hat{\boldsymbol{\Sigma}}_2 = (\mathbf{X}\mathbf{V}_n^{-1}\mathbf{X} - \mathbf{1}^t\mathbf{V}_n^{-1}\mathbf{1}\hat{\mu}_X^t\hat{\mu}_X)/n$ converges almost surely to $\boldsymbol{\Sigma}$. Since $\pi_m$ is assumed smooth and bounded, we can apply Theorem 1 from Kass, Tierney and Kadane (1990) again, and $f_n(\beta_0)$ is bounded by $(2\det\hat{\boldsymbol{\Sigma}}_2)^{-1/2}(2\pi\hat{\sigma}^2)^{(k+1)/2} \times \pi_m(\hat{\beta}_1, \hat{\sigma})$ which is a convergent quantity. Therefore, $f_n$ is uniformly bounded and by dominated convergence, the limit of $\int f_n\, d\beta_0$ equals the integral of the point-wise limit so that $D_n = D + o(1)$ as claimed in (1).

For model $M_1$ the proof is similar. The value $u_1$ is now $\hat{\beta}_1 - (\mathbf{X}^t\mathbf{V}_n^{-1}\mathbf{X})^{-1}((\beta_0 - \hat{\beta}_0)\mathbf{X}^t\mathbf{V}_n^{-1}\mathbf{1} + (\beta_{\text{top}} - \hat{\beta}_{\text{top}})\mathbf{X}^t\mathbf{V}_n^{-1}\mathbf{1}_{\text{top}})$. The term $a_n(\beta_0 - \hat{\beta}_0)^2$ is replaced by $\tilde{\beta}^t\mathbf{A}_n\tilde{\beta}$, where $\tilde{\beta}^t = (\beta_0 - \hat{\beta}_0, \beta_{\text{top}} - \hat{\beta}_{\text{top}})$ and $\mathbf{A}_n$ is the $2 \times 2$ symmetric matrix with diagonal elements $a_n$ and $\mathbf{1}_{\text{top}}^t\mathbf{V}^{-1}\mathbf{1}_{\text{top}} - \mathbf{1}_{\text{top}}^t\mathbf{V}^{-1}\mathbf{X}(\mathbf{X}^t\mathbf{V}^{-1}\mathbf{X})^{-1}\mathbf{X}^t\mathbf{V}^{-1}\mathbf{1}_{\text{top}}$, and off-diagonal element $\mathbf{1}^t\mathbf{V}^{-1}\mathbf{1}_{\text{top}} - \mathbf{1}^t\mathbf{V}^{-1}\mathbf{X}(\mathbf{X}^t\mathbf{V}^{-1}\mathbf{X})^{-1}\mathbf{X}^t\mathbf{V}^{-1}\mathbf{1}_{\text{top}}$. Note that, as before, elements in $\mathbf{A}_n$ are dominated by their first term, since $\mathbf{X}^t\mathbf{V}^{-1}\mathbf{X} = n\boldsymbol{\Sigma} + O(1)$ almost surely.

I show below that $\mathbf{A}_n$ converges to $\mathbf{W}^{-1}$ as defined in (2), whose elements are the limits of $\mathbf{1}^t\mathbf{V}_n^{-1}\mathbf{1}$, $\mathbf{1}_{\text{top}}^t\mathbf{V}_n^{-1}\mathbf{1}_{\text{top}}$ and $\mathbf{1}^t\mathbf{V}_n^{-1}\mathbf{1}_{\text{top}}$. The first quantity is $t_0^{-1}$, finite by assumption. The second quantity equals $\mathbf{1}^t\tilde{\mathbf{V}}^{-1}\mathbf{1}$, where $\tilde{\mathbf{V}}$ is obtained by pruning the tree from all tips not in the top subtree, so it converges and is necessarily smaller than $t_0^{-1}$. The third quantity exists because $(\mathbf{1}^t\mathbf{V}_n^{-1}\mathbf{1})^{-1}(\mathbf{1}^t\mathbf{V}_n^{-1}\mathbf{1}_{\text{top}})$ is $\hat{\mu}_{\text{top}}$, the estimated state at the root from character $\mathbf{1}_{\text{top}}$. Theorem 1 cannot be applied to show its convergence, because $\mathbf{1}_{\text{top}}$ is a nonrandom character, but convergence follows from the following facts: (a) $\hat{\mu}_{\text{top}}$ is the estimated state at the root from a tree where the top subtree is reduced to a single "top" leaf whose subtending branch length decreases when more tips are added to the top subtree, to a nonnegative limit. (b) On the reduced tree, $\hat{\mu}_{\text{top}}$ is the weight with which the top leaf contributes to ancestral state estimation. (c) This weight decreases as more tips are added outside the top subtree.  □

**Acknowledgments.** The author is very grateful to Thomas Kurtz for insightful discussions on the almost sure convergence result.

DEPARTMENTS OF STATISTICS AND OF BOTANY
UNIVERSITY OF WISCONSIN—MADISON
1300 UNIVERSITY AVENUE
MADISON, WISCONSIN 53706
USA
E-MAIL: ane@stat.wisc.edu